\documentclass[fleqn,twoside]{article}
\usepackage{espcrc2}

\usepackage{graphicx}
\usepackage[figuresright]{rotating}
\usepackage{psfig}

\newcommand{\HI}{H\,{\sc i}}
\newcommand{\HII}{H\,{\sc ii}}
\newcommand{\la}{\ifmmode\stackrel{<}{_{\sim}}\else$\stackrel{<}{_{\sim}}$\fi}
\newcommand{\ga}{\ifmmode\stackrel{>}{_{\sim}}\else$\stackrel{>}{_{\sim}}$\fi}

\title{Observations of Magnetic Fields in the Milky Way and in Nearby Galaxies
with a Square Kilometre Array}

\author{R. Beck\address{Max-Planck-Institut f\"ur Radioastronomie,
Auf dem H\"ugel 69, 53121 Bonn, Germany} and
B. M. Gaensler\address{Harvard-Smithsonian Center for Astrophysics,
60 Garden Street MS-6, Cambridge, MA 02138, USA}}

\begin{document}

\begin{abstract}

The role of magnetic fields in the dynamical evolution of galaxies and of
the interstellar medium (ISM) is not well understood, mainly because such
fields are difficult to directly observe.  Radio astronomy provides the
best tools to measure magnetic fields: synchrotron radiation traces fields
illuminated by cosmic-ray electrons, while Faraday rotation and Zeeman
splitting allow us to detect fields in all kinds of astronomical plasmas,
from lowest to highest densities. Here we describe how fundamental new
advances in studying magnetic fields, both in our own Milky Way and in
other nearby galaxies, can be made through observations with the proposed
Square Kilometre Array. Underpinning much of what we propose is an {\em
all-sky survey of Faraday rotation}, in which we will accumulate tens
of millions of rotation measure measurements toward background radio
sources.  This will provide a unique database for studying magnetic
fields in individual Galactic supernova remnants and \HII\ regions,
for characterizing the overall magnetic geometry of our Galaxy's disk
and halo, and for understanding the structure and evolution of magnetic
fields in galaxies. Also of considerable interest will be the mapping
of diffuse polarized emission from the Milky Way in many narrow bands
over a wide frequency range. This will allow us to carry out {\em Faraday
tomography}\ of the Galaxy, yielding a high-resolution three-dimensional
picture of the magnetic field within a few kpc of the Sun, and allowing
us to understand its coupling to the other components of the ISM. Finally,
direct synchrotron imaging of a large number of nearby galaxies, combined
with Faraday rotation data, will allow us to determine the magnetic field
structure in these sources, and to {\em test both the dynamo and primordial
field theories} for field origin and amplification.

\end{abstract}

\maketitle


\section{Introduction}

A full understanding of galactic structure and evolution is impossible
without understanding magnetic fields. Magnetic fields fill interstellar
space, contribute significantly to the total pressure of interstellar
gas, are essential for the onset of star formation, and control the
density and distribution of cosmic rays in the interstellar medium
(ISM).  However, because magnetic fields cannot be directly observed,
our understanding of their structure and origin lags significantly behind
that of the other components of the ISM.

Radio astronomy has long led the way in studying astrophysical
magnetic fields. Synchrotron emission measures the total field strength;
its polarization yields the regular field's orientation in the sky plane and
also gives the field's degree of ordering; Faraday rotation provides
a measurement of the mean direction and strength of the field along
the line of sight; the Zeeman effect provides an independent measure
of field strength in cold gas clouds. All these effects have been
effectively exploited. However, the study of magnetism in the Milky
Way and in galaxies is a field still largely limited to examination of
specific interesting regions, bright and nearby individual sources, and
gross overall structure. Here we describe how exciting new insights into
magnetic fields can be provided by the unique sensitivity, resolution
and polarimetric capabilities of the Square Kilometre Array (SKA).

\section{All-sky Rotation Measures}
\label{allsky}

\subsection{Background}

While synchrotron emission and its polarization are useful tracers of
magnetic fields, they are only easily detected in regions where the
density of cosmic rays (i.e., relativistic gas) is relatively high,
or where the magnetic field is strong. Many regions of interest for
magnetic field studies are far from sites of active star formation and
supernova activity, and thus cannot be studied through these techniques.

A much more pervasive probe of interstellar magnetic fields is {\em
Faraday rotation}, in which birefringence in the magneto-ionic ISM causes the
position angle of a linearly polarized wave to rotate.  For a wave with
emitted position angle $\phi_0$ observed at a wavelength $\lambda$,
the detected position angle is:
\begin{equation}
\phi_1 = \phi_0 + {\rm RM}~\lambda^2.
\label{eq_faraday}
\end{equation}
In this expression, the {\em rotation measure}\ (RM), in units
of rad~m$^{-2}$, is defined by :
\begin{equation}
{\rm RM}~ = K \int B \cos \theta~n_e dl,
\label{eq_rm}
\end{equation}
where $K \approx 0.81$~rad~m$^{-2}$~pc$^{-1}$~cm$^3$~$\mu$G$^{-1}$,
$B$, $\theta$ and $n_e$ are the regular magnetic field strength, inclination
of the magnetic field to the line of sight and number density of
thermal electrons, respectively, and the integral is along the line
of sight from the observer to the source.

Multiwavelength observations of polarized sources can directly yield
the RM along the line of sight.
An estimate of RM in itself does not directly yield a value for $B$,
but the sign of the RM can provide information on the {\em direction}\
of the regular magnetic field. Furthermore, in cases where something is known
about the electron density along the line of sight (e.g., from H$\alpha$
observations, thermal
radio emission, X-ray data, or pulsar dispersion measures), an
estimate of the mean amplitude of the magnetic field strength along the
line of sight can be directly inferred (a caveat is that correlations
between and clumpiness in $B$ and $n_e$ both need to be properly accounted
for in such estimates; \cite{bssw03}).

\begin{figure*}
\begin{minipage}{\textwidth}
\centerline{\psfig{file=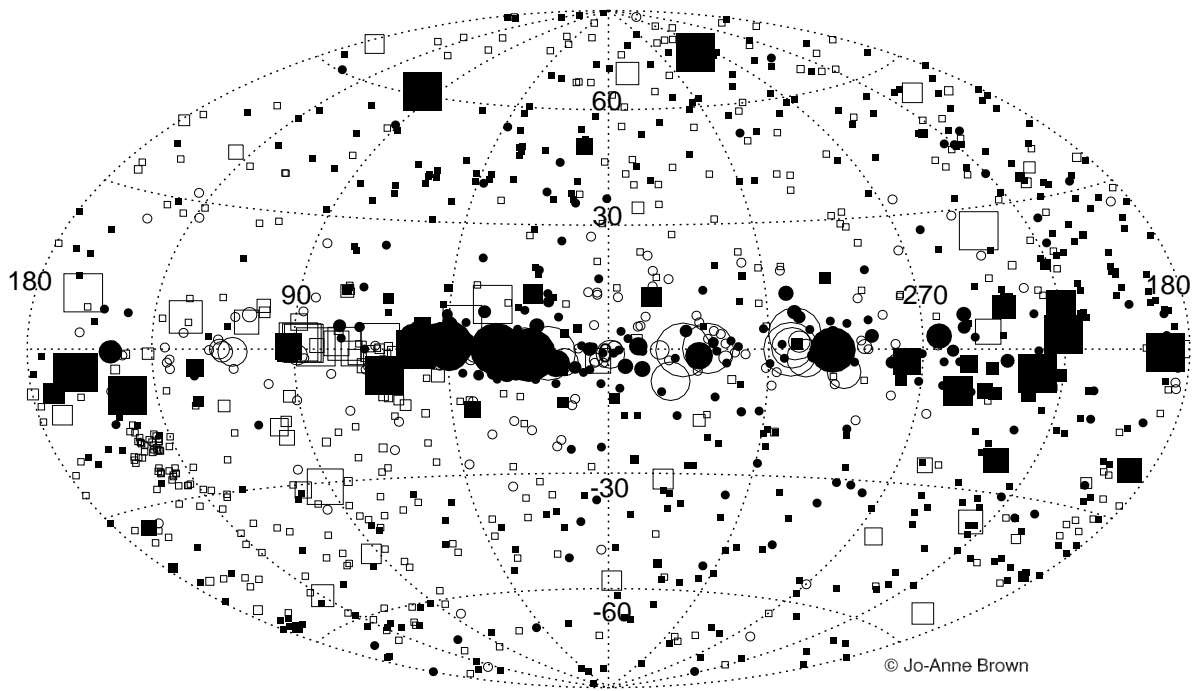,width=\textwidth,clip=}}
\caption{An Aitoff projection of the celestial sphere in Galactic
coordinates, showing a recently compiled sample of 1203 RMs.
Closed symbols represent positive RMs, while open symbols correspond
to negative RMs. In the range $100 < |{\rm RM}| < 600$~rad~m$^{-2}$, the linear
size of a symbol is proportional to $|{\rm RM}|$ for the corresponding source;
for magnitudes of RM outside this range, the sizes of symbols
are fixed at those corresponding to either $|{\rm RM}| = 100$~rad~m$^{-2}$
or $|{\rm RM}| = 600$~rad~m$^{-2}$. The 887 squares represent RMs toward
extragalactic sources, while the 316 circles indicate RMs of radio
pulsars.  Figure courtesy of Jo-Anne Brown.}
\label{fig_aitoff_rms}
\end{minipage}
\end{figure*}

Thus provided that one can find a linearly polarized source as background,
one can infer the strength and geometry of the magnetic field in
foreground material, regardless of the foreground source's synchrotron
emissivity.  However, for studying magnetic fields within the Galaxy,
a serious shortcoming of this technique is the lack of background objects
--- for both pulsars and extragalactic radio sources, limited sky coverage
and relatively poor sensitivity in polarization surveys severely limit the
number of sightlines towards which one can measure the rotation measure
in a foreground object, particularly if the source is of small angular
extent (e.g.,\ an \HII\ region or a supernova remnant).  Until recently,
there were only about 1200 RM measurements of compact sources over the
entire sky (about 900 extragalactic sources, plus 300 pulsars), as shown
in Figure~\ref{fig_aitoff_rms}. In the Galactic plane, new surveys are
expanding this sample at a rate of about one RM measurement per deg$^2$
(e.g., \cite{rbgd,btj03}); over the rest of the sky, observations
with the Effelsberg telescope of polarized NVSS sources will soon
add another $\sim1000$ RM measurements.  Such data sets are proving
useful in studying the global properties of the Galactic magnetic field
(e.g., \cite{btwm03}), but do not have a dense enough sampling for
detailed modeling or for studying discrete foreground objects
\cite{hc80,sc86}.

\subsection{A Survey for Polarized Sources with the SKA}

An exciting experiment for the SKA will be to greatly increase the
density of polarized background sources on the sky, providing enough
statistics to make possible the study of all manner of foreground
magneto-ionic sources.  We thus propose an all-sky RM survey, which can
provide a closely-packed grid of RM measurements in any direction. In
the following discussion, we assume that the SKA has a sensitivity
$A_{eff}/T_{sys} = 20\,000$~m$^2$~K$^{-1}$, of which 75\% is distributed
on baselines shorter than $\sim35$~km, that the field of view at 1.4~GHz is
1~deg$^2$, and that the total bandwidth
available at 1.4~GHz is 25\% (i.e., 350~MHz).
If we aim to survey 10\,000~deg$^2$ of the sky
in one year of integration time, the integration time per pointing is
$\sim1$~hour, and the expected sensitivity in linear polarization
(i.e., from the combination in quadrature of Stokes $Q$ and $U$)
is then $\sigma
\approx 0.1$~$\mu$Jy. While a year of observing time is a large request,
we note that such a survey has many other applications (e.g.,
Stokes~$I$ imaging, \HI\ absorption, pulsar surveys), and that it is
reasonable to assume that multiple projects could piggyback on the same
set of observations.

An important requirement for the SKA will be spectropolarimetric
capability, wherein full Stokes products will be available in multiple
contiguous channels across a broad continuum bandwidth. Provided that at least
$\sim4$ channels are available across the band, the individual channel widths
determine the maximum value of $|{\rm RM}|$ which can be measured, while
the total bandwidth determines the accuracy of the RM measurement.  For an
observing wavelength $\lambda$, a total bandwidth (in wavelength units)
$\Delta \lambda$, and a source detected with signal to noise $\mathcal{L}$
in linear polarization, the expected precision of an RM measurement is:
\begin{equation}
\Delta{\rm RM} \approx \frac{1}{2\mathcal{L}} \frac{1}{\lambda\Delta\lambda}.
\end{equation}
An appropriate observing wavelength for detecting large numbers
of RMs is $\lambda = 21$~cm: this provides a large field of view,
without introducing severe internal depolarization effects which
will prevent RMs from being measured in many extragalactic background
sources. At 21~cm, our assumed fractional bandwidth of 25\% corresponds
to $\Delta \lambda = 0.053$.  For a precision in RM of $\Delta {\rm
RM}$~$\approx 5$~rad~m$^{-2}$, we thus require a signal to noise in
polarization $\mathcal{L} \sim 10$.  (Such a precision in RM is more
than sufficient for the purposes considered here; higher precision
measurements are possible, but require correction for the effects of
the Earth's ionosphere.)

Given the high density of polarized sources expected on the sky (see below), we
require an angular resolution $\la1''$ to ensure that the polarized sky is
not confusion limited. To carry out an efficient survey, we will need to
image the full 1-deg$^2$ field of view of the SKA at this resolution. If
this field corresponds to the primary beam of a single element, then
individual channel widths need to be smaller than $\sim100$~kHz to avoid
bandwidth smearing (i.e., $>3500$~channels will
be needed across the 25\% observing
bandwidth).  This also is more than sufficient to completely mitigate
bandwidth depolarization provided $|{\rm RM}| \la 10^4$~rad~m$^{-2}$,
which is likely to be the case for almost all sightlines.  Low values of
$|{\rm RM}|$ can be directly identified from images of linear polarization
using just a few ($\sim$10) channels across the observing bandwidth and
fitting directly to the position angle swing across this band \cite{rbgd};
the polarized signal from high-RM sources will require Fourier analysis
of the polarimetric signal to identify \cite{agb96,kfze04}, but ultimately
can be recovered with the same signal-to-noise as for the low-RM case.

We conclude our description of this experiment by noting its implications
for polarization purity and for imaging dynamic range.  As discussed
in the following section, the peak in the distribution of fractional
polarization for extragalactic sources is expected to be at $\log_{10}
\Pi \approx -1.5$ (where $\Pi$ is the fractional linear polarization),
but with a significant fraction of the population extending down to
$\log_{10} \Pi \la -2.5$.
Thus in order to be able to detect polarization in the bulk of sources,
the final calibrated mosaiced images require a polarization purity of at
least --25~dB over the entire field. We assume that the final images
will incorporate averaging of overlapping fields and wide parallactic
angle coverage, which will assist in achieving this target. We note
that a higher polarization purity, closer to --40~dB, is required for
single-pointing on-axis observations, for which experiments involving
very weakly polarized sources may be important.

The density of polarized sources shown in Figure~\ref{fig_rm_grid}
implies that in any 1-deg$^2$ field, we always expect to detect at least
one source whose linearly polarized flux density is $>6$~mJy. Thus to
detect polarized sources as faint as 1~$\mu$Jy (the 10-$\sigma$ detection
limit of a 1-hour integration) a modest dynamic range of at least 6000
is required. We note also that in this same field, we expect to detect
at least one source whose {\em total}\ intensity is $\la200$~mJy. For
a polarization purity no worse than --25~dB anywhere in the field, this
will result in leakage into the polarized image at the level of 0.6~mJy;
this should not present any challenges to obtaining high dynamic range
in the polarized images.

\subsection{Detection Statistics for the SKA RM Survey}
\label{allsky_stat}

We first consider the likely detection statistics for pulsars.
The polarimetry statistics of Gould \& Lyne \cite{gl88} suggest that
pulsars have a typical linearly polarized fraction of $\Pi_{PSR} \sim
20$\%. Thus the faintest pulsar from which we can extract a useful
RM measurement will have a total intensity flux density of $\sigma
\mathcal{L}/\Pi_{PSR} \approx 5$~$\mu$Jy. This is sufficient to detect
virtually every radio pulsar in the Galaxy which is beaming towards us
\cite{kra03}, i.e.\ about 20\,000 pulsars.  We assume that most of these
pulsars lie at Galactic latitudes $|b|\le5^\circ$, so that the area of
the sky under consideration is 3600~deg$^2$. Thus the expected source
density of RMs from pulsars is about 6~deg$^{-2}$, or an angular spacing
of about one source every $30'$ in the Galactic plane.

For extragalactic background sources, we can estimate the likely
distribution of polarized sources by convolving the differential
source count distribution, $dN/dS$, by the probability distribution of
fractional polarized intensity, $P(\Pi_{EG})$. $dN/dS$ can be obtained
directly from deep continuum surveys and extrapolations thereof
\cite{hwce00,hac+03,smg04}, while $P(\Pi_{EG})$ can be estimated from
the 1.4~GHz NVSS catalog \cite{ccg+98}. Considering only NVSS sources
with flux densities $>80$~mJy so as to eliminate most sources whose
polarized fraction is below the sensitivity of the survey, we find
that $P(\Pi_{EG})$ can be fitted by two Gaussian components, centered
at $\log_{10} \Pi \approx -1.5$.  In the following discussion we
assume that this distribution of $P(\Pi_{EG})$ does not evolve as a
function of flux density, although we note that several recent studies
have suggested that weaker sources are more highly polarized
\cite{mbd+02,tmt+03,tmt+04}.  If this effect is real, our source count
estimates should be considered a lower limit.

\begin{figure*}
\begin{minipage}{\textwidth}
\centerline{\psfig{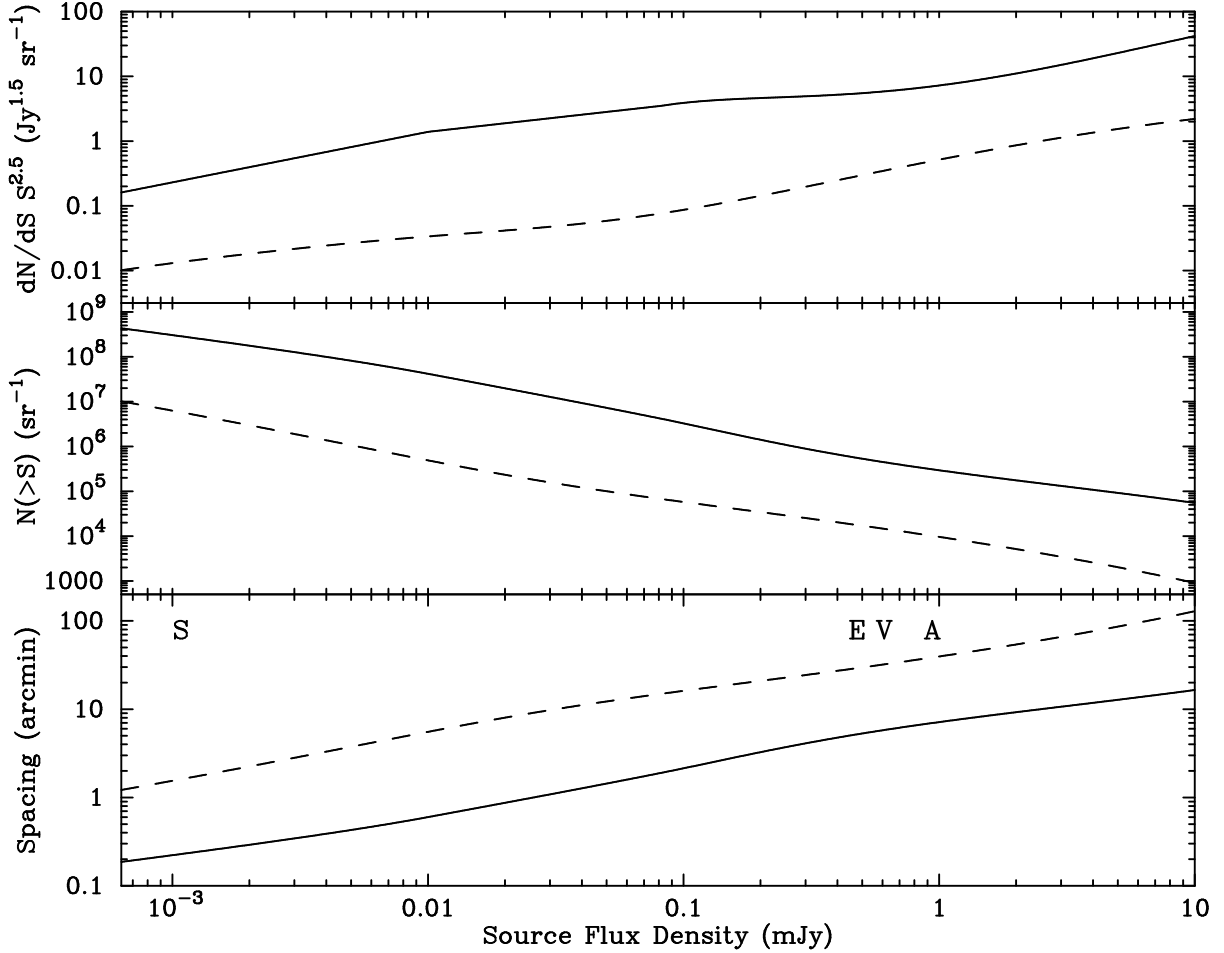}}
\caption{Distribution of extragalactic source counts (Euclidean normalized)
in both total
intensity (solid line) and in linear polarization (dashed line) at an
observing frequency of 1.4~GHz.  The upper panel shows the differential
source count distribution.  The total intensity source counts have
been calculated using the polynomial solution of Hopkins et al.\
\cite{hac+03} down to a flux density of 80~$\mu$Jy, a straight line
of slope 0.44 between 10 and 80~$\mu$Jy, and a straight line of slope
0.78 below 10~$\mu$Jy; the latter two components come from a fit to the
distribution calculated by Seymour et al.\ \cite{smg04}.  The linearly
polarized source counts are derived by convolving the total intensity
distribution by the probability distribution of fractional polarized
intensity as determined from the NVSS catalog (see \S\ref{allsky_stat}
for details). The middle panel shows the expected source density above
a given flux threshold. Both curves have been obtained by integrating
the functions shown in the upper panel; the polarized source density
corresponds only to sources for which RMs can be reliably determined,
i.e.\ about 50\% of all linearly polarized sources. The lower panel
shows the corresponding mean spacing between all background sources
(solid line) and between RM measurements (dashed line),
 as a function of minimum detectable flux.
The symbols ``A'', ``V'', ``E'' and ``S'' show representative thresholds
of linearly polarized flux density sufficient to accurately
measure RMs,
for a 20-min observation with the ATCA (total effective bandwidth $\Delta
\nu \approx 100$~MHz), a 10-min observation with the VLA ($\Delta \nu
\approx 37$~MHz), a 3-min observation with the EVLA ($\Delta \nu \approx
500$~MHz), and a 1-hour observation with the SKA ($\Delta \nu \approx
350$~MHz), respectively.}
\label{fig_rm_grid}
\end{minipage}
\end{figure*}

Convolving $dN/dS$ with $P(\Pi_{EG})$ results in a predicted
linearly polarized differential source count distribution as shown in
Figure~\ref{fig_rm_grid}.  Down to a flux limit of $\sigma\mathcal{L}
\approx 1$~$\mu$Jy, we thus expect a density of polarized sources
of $\sim2900$~deg$^{-2}$.  Only about 50\% of these sources will
have measurable RMs, usually due to internal depolarization which
destroys the RM~$\propto \lambda^2$ dependence predicted from
Equation~(\ref{eq_faraday}) \cite{btj03}.  We thus expect to find
$\sim2\times10^7$ RMs over the survey (about one RM per second of
observing time!), at a mean separation of $\sim90''$ between adjacent
measurements.  For particular regions of interest (e.g., towards a
specific supernova remnant or nearby galaxy), one could carry out a much
deeper integration to further improve the spacing of this ``RM grid''.
For example, in a 10-hour targeted observation, the mean spacing of RM
measurements would shrink to $\approx40''$.

\section{Scientific Applications of an RM Grid}

The densely-spaced RM grid which would result from the experiment
proposed above will have numerous applications: the magnetic
properties of any extended foreground object will be able to
be mapped in detail. Before discussing specific projects,
we make a few general comments about such analyses:

\begin{itemize}
\item The RM signature of an extended foreground object can only ever be
identified provided that its contribution to the total RM dominates the
average intrinsic RM of the background sources. Since the intrinsic
RM of extragalactic sources, averaged over many such objects,
is $|{\mbox{\rm RM}}| \la 5$~rad~m$^{-2}$,
we should be easily able to identify the extra RM signal produced by
intervening supernova remnants (SNRs) and \HII\ regions, or by  diffuse
magnetic fields in the Milky Way, in other galaxies or in galaxy clusters.  \\

\item A RM only probes the line-of-sight component of an object's regular
magnetic field. For studies of individual objects, full three-dimensional
geometries can be inferred by combining such measurements with other
probes: e.g., linear polarization position angles of synchrotron emission
for SNRs, infrared polarimetry for \HII\ regions, or depolarization
effects (e.g., \cite{fbbs04}). However, for studies of an ensemble of
objects, or of turbulent processes, the RM measurements alone can suffice.
We also note that
thanks to our position within the Milky Way, the three-dimensional
field geometry can be inferred by considering RM
properties in different parts of the sky, since each RM
probes the parallel field component in a different direction.
This method is also applicable to nearby galaxies which are significantly
inclined with respect to the line of sight \cite{fbbs04}. \\

\item While pulsars have a much lower source density than extragalactic
sources, all pulsars will also have measurements of their dispersion
measure (DM). Dividing the RM by the DM can allow a direct estimate of
the mean magnetic field strength along the line of sight, subject to
caveats on possible correlations between the magnetic field strength and
the electron density \cite{bssw03}. For nearby pulsars, distances will
be directly available from parallax measurements \cite{ccv+04}; for more
distant sources the distance can be estimated from the DM \cite{cl02}.

\end{itemize}

\subsection{The Milky Way}

Our own Galaxy provides a wonderful opportunity for a detailed study of
the generation and amplification of galactic magnetic fields, and their
coupling to the ISM.  However, we currently lack a good understanding
of the overall field geometry, including the number and location of
reversals, the pitch angle of the presumed spiral pattern of the disk's
magnetic field, and the structure of the magnetic field in the halo
\cite{rbbe3}.  This results both from the sparse sampling
of RMs, and from the lack of sensitivity to pulsars in the more distant
parts of the Galaxy.

\begin{figure*}
\begin{minipage}{\textwidth}
\centerline{
\psfig{file=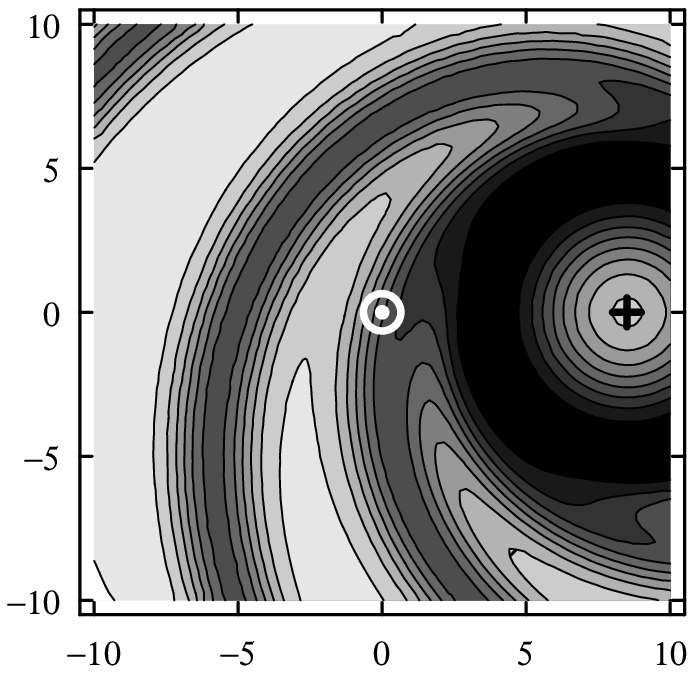,width=0.333\textwidth}
\psfig{file=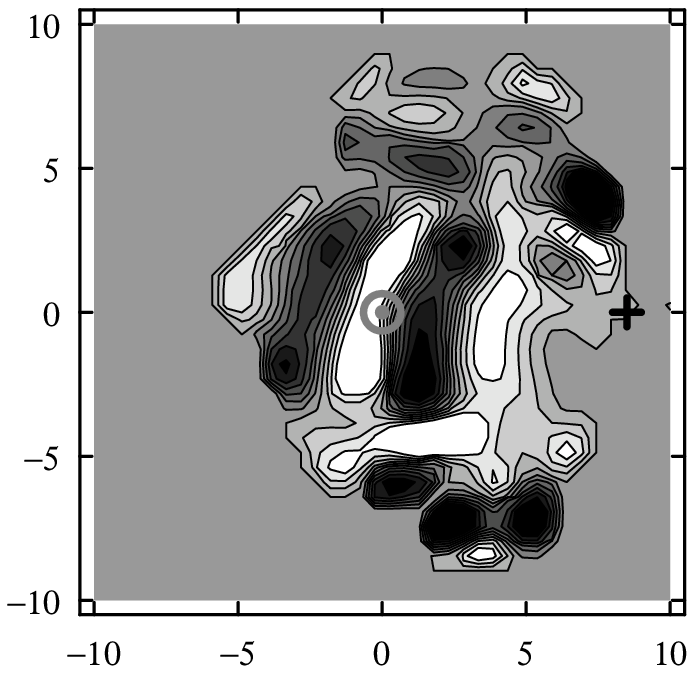,width=0.333\textwidth}
\psfig{file=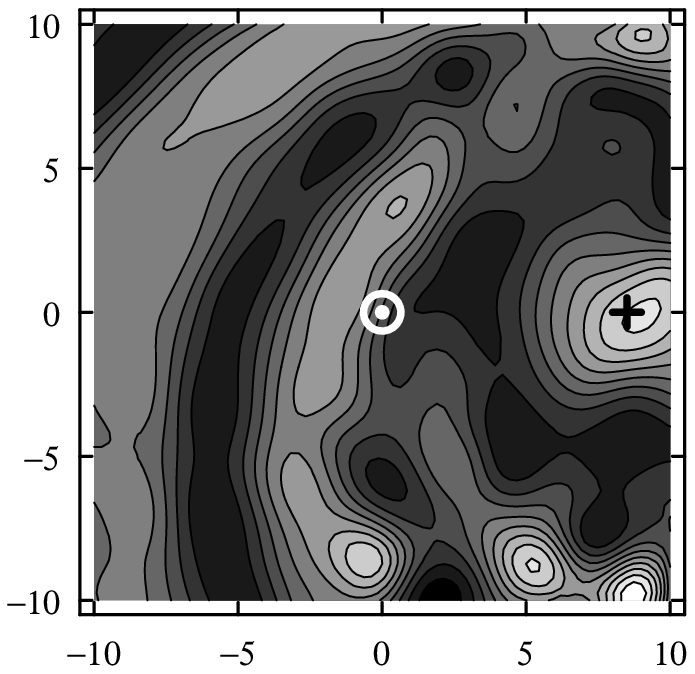,width=0.333\textwidth}}
\caption{Demonstration of how the structure of the
Galactic magnetic field can be inferred
from pulsar RMs using wavelet transforms \cite{sfss02}.
Left panel: A simple model for the Galactic magnetic field.
The axes are in units of kpc, and the position of the Sun
is marked by the white circle. Darker regions represent
regions of higher magnetic field. Center panel: the inferred
magnetic field structure using the $\sim300$ pulsar RMs
listed in the catalog of Taylor et al.\  \cite{tml93}.
Right panel: the inferred magnetic field structure
determined from a hypothetical distribution of 1600 randomly distributed
pulsars. With 20\,000 pulsar RMs, the SKA will be able to fully
delineate the global magnetic field structure of the Galaxy.
Reproduced with permission from ESO.}
\label{fig_wavelet}
\end{minipage}
\end{figure*}

The SKA RM survey proposed above will overcome all these
limitations. Using wavelet transforms, the Galactic pulsar RM distribution
can be directly inverted to delineate the full geometry of the Galactic
magnetic field in the disk (Fig.\ \ref{fig_wavelet}).  Extragalactic
RMs can trace components of the magnetic field such as the outer arms
\cite{btwm03} and the vertical structure of the field in the disk and
halo \cite{hmbb97}, and can also identify loops and other features which
trace Galactic structure (e.g., \cite{sk80}).

The RM grid will also be a powerful probe of turbulence.  Turbulence is
thought to be injected at specific size scales into the ISM by various
processes, and then cascades down to increasingly smaller scales
before ultimately being dissipated as heat. It has been argued that
the amplitudes of turbulent fluctuations in the ISM form a single power
spectrum extending from the largest Galactic scales ($\ga$~1~kpc) down to
a tiny fraction of an AU \cite{ars95}.  However, there are observational
indications that the slope of the power spectrum may deviate from that
predicted by the standard three-dimensional Kolmogorov model for the
cascade (e.g., \cite{ms96,rbsh}). Futhermore, it is likely that the
turbulent properties of the ISM strongly depend on location within the
Galaxy, as might be expected if SNRs or \HII\ regions are responsible
for providing much of the energy which goes into turbulent motions
(e.g., \cite{rbhg}).  Finally, our understanding of how fluctuations
in magnetic field strength couple to those in electron density is still
limited (e.g., \cite{mg01,cl03}).

An ensemble of polarized extragalactic background sources can be used
to compute structure functions, wavelet transforms and autocorrelation
functions of RM (e.g., \cite{lsc90,ms96,fsss01,ev03}), all of which
provide different information on the combined spatial power spectrum
of density and magnetic field fluctuations in ionised gas.  However,
the sparse sampling of current data sets prevents a detailed analysis
on scales much smaller than a degree. With the SKA, the dense grid of RM
measurements can allow computation of virtually continuous power spectra
of RM, on scales ranging from $\ll1'$ up to tens of degrees. This will
allow a full characterization of magneto-ionic turbulence in this range
of angular scales.\footnote{SKA experiments relating to ISM turbulence on
much smaller scales are discussed in Lazio et al., this volume.}
We will also be able to generate
such power spectra as a function of Galactic longitude and latitude, so as
to establish how the turbulent properties of the ionised ISM vary, e.g.,
between arm and inter-arm regions, between star-forming and quiescent
regions, or between the thin disk, the thick disk and the halo.

\begin{figure*}
\begin{minipage}{\textwidth}
\centerline{\psfig{figure=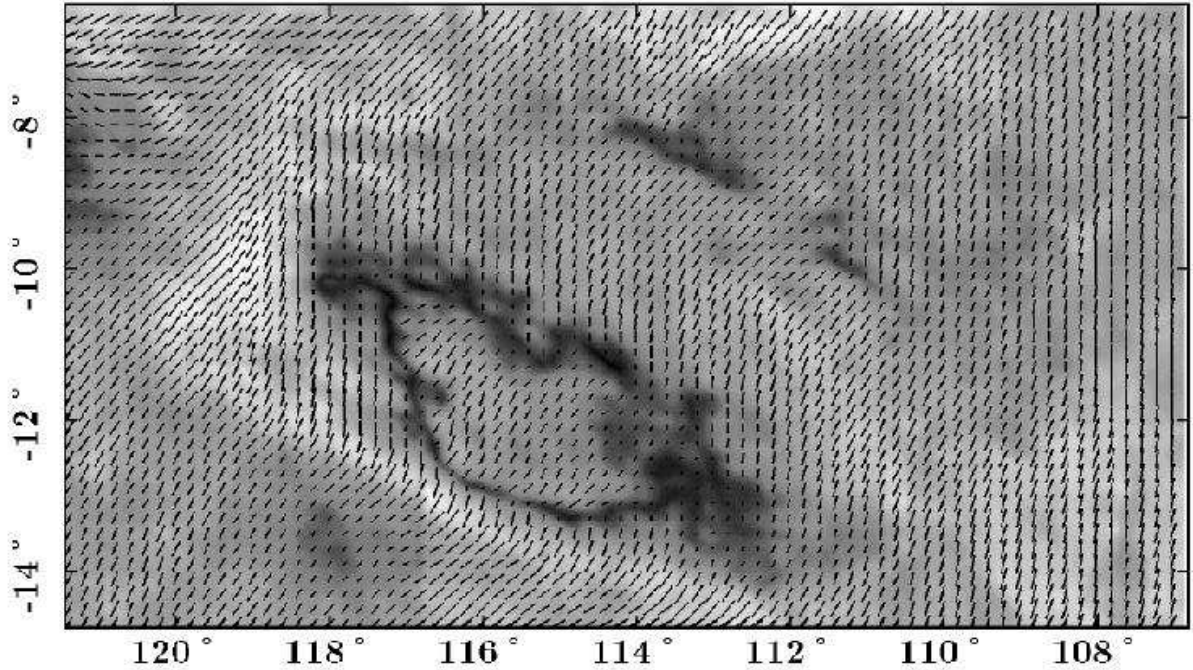,angle=270,width=0.99\textwidth,clip=}}
\caption{Linearly polarized emission and electric field vectors at
1.4~GHz from a region in Cassiopeia, possibly corresponding to a large
``magnetic bubble'' \cite{rbrf}.  Coordinates are Galactic longitude and
latitude. The image is a combination of high-resolution observations
taken with the Effelsberg 100-m telescope and large-scale emission
provided by data from the Dwingeloo telescope.}
\label{rbgal}
\end{minipage}
\end{figure*}

An additional resource will be the diffuse polarization seen all over
the sky, such  as is shown in Figure~\ref{rbgal}.  In such fields
the RM can be measured at every pixel, allowing one to compute power
spectra at even higher spatial resolutions than made possible by the RM
grid of background sources \cite{ev03,rbhg}.  Such analyses are highly
complementary to those toward the RM grid, because they probe a smaller
sightline through the Galaxy, and thus give information on structure
in specific local regions (see further discussion in \S\ref{tomo}).

At high latitudes, where optical extinction is minimal, the RMs of
background sources can be directly compared to the H$\alpha$ emission
from diffuse ionised gas and to the DMs of pulsars in globular clusters,
to separately solve for the mean electron density, gas clumping factor,
magnetic field strength and direction, and the fluctuations in each of
these quantities.

\subsection{Galactic Supernova Remnants}

Magnetic fields in SNRs are a key diagnostic of the physical processes
which govern heating, turbulence and particle acceleration generated
by strong shocks. Magnetic field strengths in SNR shocks are typically
thought to be at the level of $\la$~1mG \cite{fr04}, but it is unclear how
these high magnetic fields are produced, especially in young adiabatic
SNRs where the compression ratio is small. Turbulent amplification
in Rayleigh-Taylor unstable regions between the forward and reverse
shocks of the SNR can generate strong magnetic fields \cite{jn96a,jn96b}.
This may in turn result in significant second-order Fermi acceleration of
cosmic rays, and could have an important bearing on the overall evolution
of SNRs \cite{ost99,gw03}.

With current data we are unable to address these issues.  While the
position angle of polarized emission provides the {\em orientation}\
of the magnetic field \cite{mil90,fr04}, we normally do not have a
good estimate of the field {\em strength}\ in these regions, because
there is little reason to assume that magnetic fields and cosmic rays
are in equipartition.  Only in a few sources can we directly infer the
magnetic field strength via Zeeman splitting of shock-excited OH masers
\cite{bfgt00}, but these are likely to be special cases where the shock
is radiative and is interacting with a molecular cloud \cite{che99}.
Once again, the RM grid provides the potential to directly measure
magnetic fields on small scales in these sources.  Specifically, one
can combine RM measurements with observations of thermal X-rays from SNR
shocks, to separate out density and magnetic field contributions to the
RM \cite{mldg84}. With the SKA, this technique can be applied to many
SNRs, embedded in many different environments, and at a wide range of
evolutionary stages.

A related experiment corresponds to the effect that SNRs should have on
their surroundings.  It is widely believed that SNRs accelerate cosmic
rays through diffusive shock acceleration. An important part of this
process is that particles streaming away from the shock generate enhanced
magnetohydrodynamic turbulence just upstream, which in turn provides
the scattering centers which reflect particles back across the shock
(e.g., \cite{abr94}).  Because of this process, it is reasonable to
suppose that SNR shocks inject significant amounts of turbulent energy
into their surroundings, which ultimately become part of the overall
turbulent cascade. These physical processes can be directly tested
by observations of RMs of background sources, in that we expect to see enhanced
amplitudes and dispersions of RMs immediately beyond the bright radio
rims of young SNRs. A preliminary effort in this regard tentatively
suggests that indeed there is a larger scatter in the RMs of sources behind
SNRs than in other regions \cite{sim92}, but from the crude statistics
of that study it is difficult to disentangle the effect of a SNR itself
from the complicated environment in which it is often embedded. The much
denser sampling which the SKA will generate, accumulated over many SNRs
in the Galactic plane, can provide a definitive study of this effect.

Finally, it is thought that in many cases (most notably around the Crab
Nebula), the SNR shock is invisible, perhaps because it has not yet
interacted significantly with the ISM (e.g., \cite{fkcg95}). Such shocks
might be detectable by their effect on the RMs of background sources;
such a detection
would greatly add to our understanding of the evolution of young SNRs
expanding into low density regions or into stellar wind bubbles.

\subsection{Galactic \HII\ Regions}

\HII\ regions provide a link between the molecular clouds from which stars
form, the powerful winds which massive stars generate, and the ambient
ISM into which all this material ultimately diffuses. Because \HII\
regions span a very wide range of densities, studies of magnetic fields
in \HII\ regions provide an insight into how magnetic fields control the
flow of gas, and conversely how compression of gas can amplify magnetic
fields \cite{th86}. Zeeman splitting of masers can probe magnetic fields
in ultracompact \HII\ regions, but it is difficult to measure magnetic
fields in more diffuse sources. Heiles \& Chu \cite{hc80} demonstrated
that RMs of background sources can yield such measurements, but until now this
technique has had limited application because of the sparseness of
such background sources. With the SKA, the RM produced by \HII\ regions
can be probed in detail using such background sources; since electron
densities are readily determined from H$\alpha$ or radio continuum
observations, the magnetic field strength can be easily extracted from
the RM \cite{rbgd}. With these data we can characterize how gas and
magnetic fields are compressed by ionization fronts, how the ionization
fraction within photo-dissociation regions varies with distance from
the central star, and what role magnetism plays in the highly turbulent
interiors of \HII\ regions.

Finally, we note that since RM signals from low density ionized regions are
much more easily identified than emission measures or other tracers,
many groups are now identifying unusual regions of ionised gas which are
only seen by their effect on
the diffuse polarized emission of the Galactic background (Fig.\ \ref{rbgal};
\cite{gldt98,rbhk2}). With the wide field of view of the SKA, we expect
to be able to identify many more such sources. The continuous frequency
coverage of these observations  will allow detailed tomographic studies
(see \S\ref{tomo}), which can allow us to better establish these
sources' basic properties.

\subsection{Nearby Galaxies}

Magnetic fields can be detected via synchrotron emission only if there
are cosmic-ray electrons to illuminate them. Cosmic rays are probably
accelerated in objects related to star formation. However, the radial
scale length of synchrotron emission in nearby galaxies is much larger
than that of the star formation indicators like infrared or CO line
emission \cite{rbbe4}. Magnetic fields must extend to very large radii,
much beyond the star-forming disk.  In the outermost parts of galaxies
the magnetic field energy density may even reach the level of global
rotational gas motion and affect the rotation curve \cite{rbbf}.

\begin{figure*}
\begin{minipage}{\textwidth}
\centerline{\psfig{figure=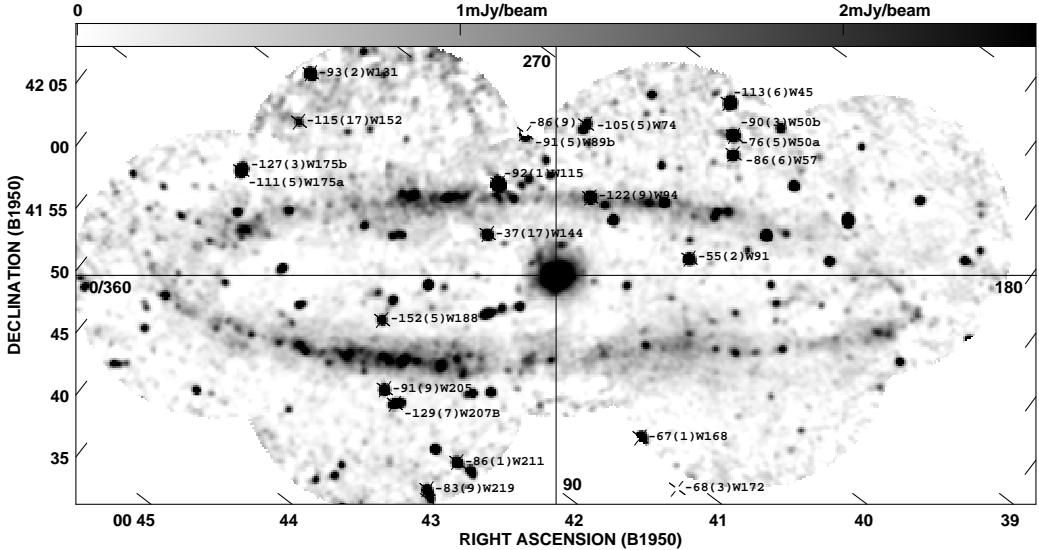,angle=270,width=14cm}}
\caption{Polarized radio sources behind M~31, superimposed onto
the 1.4~GHz continuum map derived from the VLA and Effelsberg
data \cite{rbhbb}. The numbers are RM values (between 1.4 and 1.7 ~GHz)
with standard deviations in brackets, followed by the 37W catalog names.}
\label{rbm31a}
\end{minipage}
\end{figure*}

Field strengths in the {\em outer}\ parts of galaxies can only be measured
by Faraday rotation measures of polarized background sources.  Han et
al.\ \cite{rbhbb} found evidence for regular fields in M~31 at 25~kpc
radius of similar strength and structure as in the inner disk.  However,
even within the huge field of M~31 observed with the VLA (B array)
at 1.4 and 1.7~GHz, only 21 sources with sufficient polarization flux
densities were available (Fig.~\ref{rbm31a}).  With only a few detectable
polarized sources per square degree at current sensitivities (see Fig.\
\ref{fig_rm_grid}), no galaxies beyond M~31 can be mapped in this way.

The SKA will dramatically improve the situation. Within the fields of
M~31, the LMC or the SMC (a few square degrees each), a deep observation
could provide $>10^5$ polarized background sources, and thus allow
fantastically detailed maps of the magnetic structure. The field of a
spiral galaxy at 10 Mpc distance will still include about 50 sources;
several hundred galaxies could be studied in this way.

The sensitivity of the rotation measure maps obtained by smoothing
of the RM grid will be better than 1~rad~m$^{-2}$, allowing us to
detect fields weaker than $\sim1$~$\mu$G in a halo of $10^{-3}$~cm$^{-3}$
electron density, or ionised gas of less than $2\times10^{-4}$~cm$^{-3}$
electron density in a 5~$\mu$G regular field (assuming a pathlength
of 1~kpc for both cases).  The SKA will be by far the most sensitive
detector of magnetic fields and ionised gas in the outskirts of galaxies
and in the intergalactic medium.

\section{Faraday Tomography}
\label{tomo}

Major progress in detecting small structures has recently been
achieved with decimetre-wave polarization observations in the Milky Way
\cite{rbdh,rbdr,rbgd,rbhk1,rbhk2,rbrf,rbuf1,rbuf2,rbul}.  A wealth of
structures on parsec scales has been discovered: filaments, canals,
lenses, and rings (e.g., Fig.\ \ref{rbgal}).  Their common property is
that they appear only in maps of polarized intensity, but not in total
intensity.  Some of these features directly trace small-scale structures
in the magnetic fields or ionised gas. Other features are artifacts due
to Faraday rotation ({\em Faraday ghosts}) and may give us new information
about the properties of the turbulent interstellar medium \cite{rbsb}.
To distinguish and interpret these phenomena, a multifrequency
approach has to be developed.

At the low frequencies of these polarization surveys, strong Faraday
depolarization occurs both in regions of regular field (differential
Faraday rotation) and in those of random fields (Faraday dispersion)
\cite{rbsbs}. The effect of Faraday depolarization is that the ISM
is not transparent to polarized radio waves; the opacity varies
strongly with frequency and position on sky.  For Faraday dispersion
the observable depth can be described by an exponentially decreasing
function. At low frequencies only emission from nearby regions can be
detected. In the local ISM, the typical observation depth is $\ga5$~kpc
around 5~GHz, 1--5~kpc around 1~GHz, and $\la500$~pc around 0.3~GHz.
Different frequencies trace different layers of polarized emission;
we call this new method {\em Faraday tomography}.  Individual field
structures along the line of sight can be isolated, either by their
polarized synchrotron emission, by Faraday rotation or by Faraday
depolarization of the
diffuse emission from the Galactic background.  The distances
to these polarized structures can be determined by measuring
the \HI\ absorption to these features in Stokes $Q$ and $U$ \cite{dic97}.
With a large number of channels, ``RM synthesis'' becomes possible
\cite{agb96,kfze04}, where the channel width determines the observable RM range
and the total frequency range determines the half-power width of the
``RM visibility function''. This method also allows us to distinguish
emitting regions at different distances along the same line of sight
through their different RMs.
This tomography database can be used to study the properties of magneto-ionic
turbulence in the ISM by calculating the structure function or wavelet
transforms of the three-dimensional structure in polarized intensity
and RM (a few two-dimensional studies of limited areas have already
been carried out \cite{rbsh,rbhg}).

\begin{figure}
\centerline{\psfig{figure=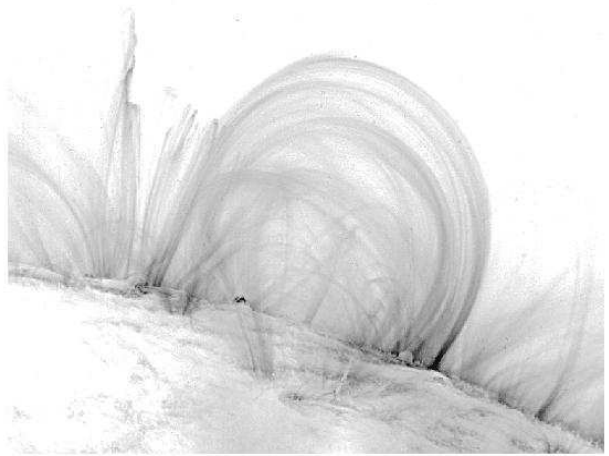,width=0.5\textwidth,clip=}}
\caption{X-ray image of the Sun, observed
by the {\em TRACE}\ satellite on 1999 Nov 6
(reproduced from {\tt http://vestige.lmsal.com/TRACE/POD/
TRACEpodarchive.html}).
}
\label{rbsun}
\end{figure}

Present-day radio polarization observations reveal only unsharp images
of magnetic fields in the ISM. The high resolution of the SKA will show
field structures illuminating the dynamical interplay of cosmic forces
\cite{rbpa}: loops, twisted fibres, and field reversals, as observed on
the Sun (Fig.~\ref{rbsun}).  The widths of such structures can be in the
range 0.1--1~pc, but are probably larger in irregular and dwarf galaxies.
With a regular field strength of 30~$\mu$G (in equipartition with cosmic rays)
and 1~pc extent along the line of sight, a (distance-independent)
polarization surface brightness at 5~GHz of 0.2~$\mu$Jy per $1''$ beam is
expected.  The SKA will be able to detect such features in the Magellanic
Clouds ($1''$ = 0.24~pc) and in M~31 / M~33 ($1''$ = 3.5~pc) within a few
hours. In the Milky Way they are hard to detect among many other emitters
along the line of sight, but Faraday tomography can isolate them.

Bright synchrotron filaments have been detected near the Galactic
Centre with milligauss field strengths \cite{rbmm}. In the ``Arc''
and the ``Snake'', particle acceleration probably occurs in reconnection
regions \cite{rbbl,rblr}. Magnetic reconnection may be a common
process in the ISM and an important heating source also in galaxy
halos \cite{rbbln}, but only the most prominent regions in the Milky Way
are visible with present-day telescopes. Acceleration of cosmic rays
should produce strong synchrotron emission in a small volume.
Even a relatively weak regular field of strength 50~$\mu$G with 1~pc
extent, in equipartition with cosmic rays, generates emission
with a polarization surface brightness at 5~GHz of 1.5~$\mu$Jy per
$1''$ beam which clearly emerges above the background. Field reversals
across reconnection regions should also be detectable via rotation
measures.

The angular resolution of the SKA will allow us to trace directly how
interstellar magnetic fields are connected to gas clouds.  The close
correlation between radio continuum and mid-infrared intensities within
galaxies \cite{rbfb} indicates that a significant fraction of the magnetic
flux is connected to gas clouds. Photo-ionisation may provide sufficient
density of thermal electrons in the outer regions of gas clouds to
hold the field lines. Observational support comes from the detection
of Faraday screens in front of a molecular cloud in the Taurus complex in
our Galaxy \cite{rbwr}. As no enhanced H$\alpha$ emission
has been detected in this direction, the local field enhancement must be significant.
Faraday tomography with the SKA will allow to
detect such Faraday screens toward molecular clouds throughout the Milky Way
and in nearby galaxies.  A $1''$ SKA beam resolves 3.5~pc in M~31 where
clouds at various evolutionary stages are available.  SKA will allow to
trace these and contribute to solving the puzzle of star formation.

It is essential that the frequencies of Faraday tomography observations
are distributed continuously over a broad bandwidth.
The frequency range has to be
chosen according to the strengths of the regular and random fields,
the electron density and the pathlength through the Faraday-rotating
region. The RM grid (see \S\ref{allsky}) helps choose the best
frequencies. To cover a wide range of physical parameters, the SKA has
to operate in the frequency range of at least 0.5--10~GHz, but 0.3--20~GHz
is desirable. Sensitive imaging of extended structures requires that
most of the collecting area lies on baselines shorter than 100~km.

\section{Dynamo versus Primordial Field Origin}
\label{dynamo}

The observation of large-scale patterns in RM in
many galaxies \cite{rbbe2} proves that the regular field in galaxies has
a {\em coherent direction}\ and hence is not generated by compression or
stretching of irregular fields in gas flows.  In principle, the dynamo
mechanism is able to generate and preserve coherent magnetic fields, and
they are of appropriate spiral shape \cite{rbbb} with radially decreasing
pitch angles \cite{rbbe1}. However, the physics of dynamo action is far
from being understood and faces several theoretical problems \cite{rbbs,rbku}.
Primordial fields, on the other hand, are hard to preserve over
a galaxy's lifetime due to diffusion and reconnection
because differential rotation winds them up. Even if they survive,
they can create only specific field patterns that differ from
those observed \cite{rbshu,rbsf}.

\begin{figure*}
\begin{minipage}{\textwidth}
\psfig{figure=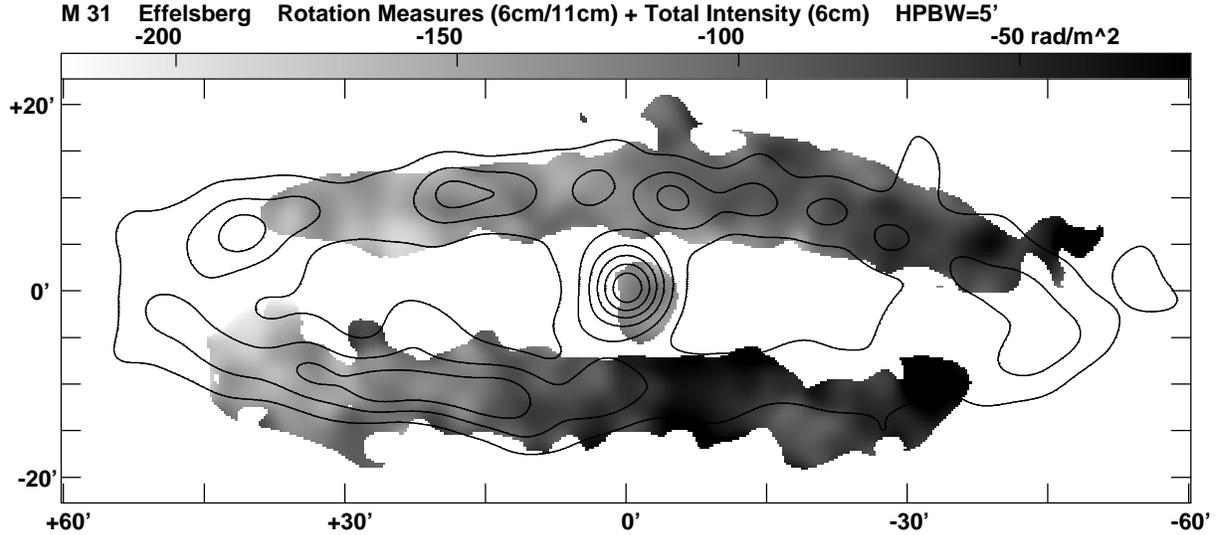,angle=270,width=\textwidth}
\caption{Faraday rotation measures in M~31 between 2.7 and 4.8~GHz,
superimposed onto the 4.8~GHz Effelsberg map \cite{rbbbh}. The rotation
measure of the Galactic foreground is about --90~rad~m$^{-2}$.}
\label{rbm31b}
\end{minipage}
\end{figure*}

The widely studied {\em mean-field $\alpha$--$\Omega$ dynamo model}\
needs differential rotation and the $\alpha$ effect (see below).
Any coherent magnetic field can be represented as a superposition
of modes of different azimuthal and vertical symmetries. The existing
dynamo models predict that several azimuthal modes can be excited
\cite{rbbb}, the strongest being $m=0$ (an axisymmetric spiral field),
followed by the weaker $m=1$ (a bisymmetric spiral field), etc.
These generate a Fourier spectrum of azimuthal RM patterns. The
axisymmetric mode with even vertical symmetry (quadrupole) is excited
most easily. Primordial field models predict bisymmetric fields or
axisymmetric fields with odd (dipole) symmetry \cite{rbsf}. For most
of about 20 nearby galaxies observed so far, the RM data indicate
a mixture of magnetic modes which cannot be reliably determined due to
low angular resolution and/or low signal-to-noise ratios \cite{rbbe2}.
M~31 is an exception with a strongly dominating
axisymmetric field (Fig.~\ref{rbm31b}).

The SKA will be able to confidently determine the Fourier spectrum of
dynamo modes. To detect an azimuthal mode of order $m$, the
spatial resolution has to be better than $\approx r/[10\times (m+1)]$
where $r$ is the galaxy's radius. With $m=4$ and $r=10$~kpc
a spatial resolution of 0.2~kpc is needed which is presently available
(with sufficient sensitivity) only for galaxies in
the Local Group.

Typical polarization intensities of nearby galaxies at 5~GHz are
$\sim$0.1~mJy per $15''$ beam. Within a $1''$ beam, 0.4~$\mu$Jy is expected,
which the SKA can detect in $\sim1$ hour of integration. Hence, the SKA can
resolve all modes up to $m=4$ in galaxies out to a distance of 40~Mpc.
The RM grid discussed in \S\ref{allsky} can identify the best candidates.

The SKA has the potential to increase the galaxy sample with well-known
field patterns by up to three orders of magnitude. The conditions for
the excitation of dynamo modes can be clarified. For example, strong
density waves are claimed to support the $m=2$ mode while companions
and interactions may enhance the bisymmetric $m=1$ mode. A dominance
of bisymmetric fields over axisymmetric ones would be in conflict with
existing dynamo models and would perhaps support the primordial
field origin \cite{rbsf}.

Dynamo models predict the preferred generation of quadrupolar patterns
in the disk where the field has the same sign above and below the plane,
while the field may be dipolar in the halo (\cite{rbbb}).
Primordial models predict dipolar patterns in the disk and in the halo
with a reversal in the plane \cite{rbsf}, which can
be distinguished using RMs in edge-on galaxies.
However, the polarized emission from radio halos is weak so that
no single determination of the vertical field symmetry has been
possible yet. This experiment also must await the SKA.

The all-sky RM grid  will allow us to trace coherent fields to large
galactic radii, well beyond the regions in which star formation takes
place, and to derive restrictions for the {\em $\alpha$ effect}. This is
an essential ingredient of dynamo action and describes the mean helicity
of turbulent gas motions.  If the $\alpha$ effect is driven by supernova
remnants or by Parker loops, dynamo modes should  be excited preferably
in the star-forming regions of a galaxy. But if the magneto-rotational
(Balbus-Hawley) instability is the source of turbulence and of the $\alpha$
effect \cite{rbsba}, magnetic field amplification with some fraction of
regular fields will be seen out to large galactic radii.

\begin{figure}
\centerline{\psfig{figure=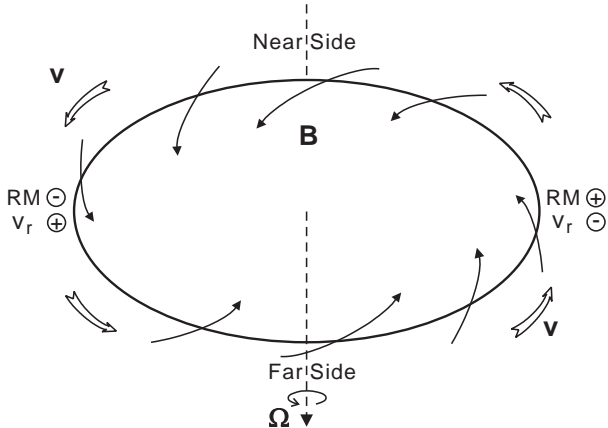,width=0.5\textwidth}}
\caption{Direction of magnetic fields in inclined galaxies with trailing
spiral arms, determined from the signs of Faraday rotation measure RM and
rotational velocity $v_r$ along the line of sight \cite{rbkb}.}
\label{rbrm}
\end{figure}

The few galaxies known to host a dominating axisymmetric $m=0$
mode possess a radial field component which is directed {\em inwards}\
everywhere \cite{rbkb}. As the field direction in dynamo models is
arbitrary, it preserves the memory of the seed field which may be
a regular primordial field with a preferred direction. The sign of the
field component along radius follows from observations of Faraday rotation
(RM) and of rotational velocity along the line of sight ($v_r$) on
both sides of the galaxy's major
axis (Fig.~\ref{rbrm}): opposite signs of RM and $v_r$ indicate
an inward-directed field, while the same signs an outward-directed
field. SKA's sensitivity will allow to observe a large galaxy sample
(a $1''$ beam will let us study such sources out to 100 Mpc) and will
clearly show any preferred field direction.

The lack of a coherent magnetic field in a resolved galaxy would indicate
that the timescale for dynamo action is longer than the galaxy's age,
or that the mean-field dynamo does not work at all.  The role of the
dynamo may still be important for the transformation of turbulent
kinetic energy into magnetic energy (the {\em fluctuation dynamo}\,
\cite{rbbla,rbsu}). Unlike the mean-field dynamo discussed above, the
fluctuation dynamo amplifies and maintains only turbulent, incoherent
magnetic fields and does not rely on overall differential rotation and
the $\alpha$ effect. This process seems to
work in all types of galaxies as long as their star formation activity
is sufficiently high. For example, dwarf irregular galaxies with almost
chaotic rotation host turbulent fields with strengths comparable to
spiral galaxies, but have no large-scale coherent fields \cite{rbck}.

\section{Summary}

A 1.4-GHz {\em all-sky survey of Faraday rotation} will accumulate tens
of millions of rotation measure measurements toward background radio
sources. This will allow us to characterize the overall magnetic geometry
and turbulent properties of the disk and halo of the Milky Way, and of
embedded individual objects such as \HII\ regions and supernova remnants.
In a highly complementary fashion, mapping of diffuse polarized emission
from the Milky Way in many narrow bands over a wide frequency range will
allow us to carry out {\em Faraday tomography}\ of the local Galaxy.
The observing frequency for this tomography needs to be tuned to the
ISM properties under study, e.g., $\sim0.5$~GHz for low-density regions
and $\sim5$~GHz for high-density regions. The combination of these
observations will yield a high-resolution three-dimensional picture of the
magnetic field within a few kpc of the Sun.  Direct synchrotron imaging
of a large number of nearby galaxies (at frequencies $>5$~GHz where
Faraday depolarization is minimal) will uncover the detailed magnetic
field structure in these sources.  Together with Faraday rotation data
from diffuse emission and from the all-sky survey of background sources,
we will be able to test both the dynamo and primordial field theories for
field origin and amplification and hence can establish an understanding
of the structure and evolution of magnetic fields in galaxies.

\vskip 0.2truein

We thank Andrew Hopkins and Nick Seymour for providing information
on source count distributions, Jo-Anne Brown for supplying
Figure~\ref{fig_aitoff_rms}, and Wolfgang Reich for Figure~\ref{rbgal}. We
also thank Elly Berkhuijsen, Marijke Haverkorn, John Dickey,
Wolfgang Reich and Anvar Shukurov for many
useful comments.  B.M.G. acknowledges the support of the National Science
Foundation through grant AST-0307358.  The {\em Transition Region and
Coronal Explorer}\ ({\em TRACE}), is a mission of the Stanford-Lockheed
Institute for Space Research, and part of the NASA Small Explorer program.
The National Radio Astronomy Observatory is a facility of the National
Science Foundation operated under cooperative agreement by Associated
Universities, Inc.

\end{document}